# EBT2 Dosimetry of X-rays produced by the electron beam from PFMA-3, a Plasma Focus for medical applications

E Ceccolini[1], F Rocchi[1,2], D Mostacci[1], M Sumini[1] and A Tartari[3]

[1] Montecuccolino Nuclear Engineering Laboratory - DIENCA, University of Bologna, via dei Colli 16, I-40136 Bologna, Italy
[2] UTFISSM-PRONOC, ENEA, via Martiri di Monte Sole 4, I-40129 Bologna, Italy
[3] Department of Physics, Ferrara University, via Saragat 1, I-44122 Ferrara, Italy

E-mail: domiziano.mostacci@unibo.it

**Abstract**. The electron beam emitted from the back of Plasma Focus devices is being studied as a radiation source for IORT (IntraOperative Radiation Therapy) applications. A Plasma Focus device is being developed to this aim, to be utilized as an X-ray source. The electron beam is driven to impinge on 50 μm brass foil, where conversion X-rays are generated. Measurements with gafchromic film are performed to analyse the attenuation of the X-rays beam and to predict the dose given to the culture cell in radiobiological experiments to follow.

## 1. Introduction

Plasma Focus devices are well known to emit during their pinch phase a collimated beam of electrons in the backward direction, i.e., the direction opposite to that of macroscopic motion of the plasma sheet between the electrodes. The beam is very intense ($1.25 \cdot 10^{25}$ particles per discharge), and lasts the time of the pinch, typically a few tenths of nanoseconds (Ceccolini *et al* 2011). The beam can possibly be utilized as a source in radiation therapy, particularly in IORT (IntraOperative Radiation Therapy) applications (Sumini M *et al* 2010, Tartari *et al* 2004); due to the short delivery time and to the possibility of a high repetition rate, large doses can be delivered in a very limited time, producing a very high dose rate. Within the framework of a research project of the Alma Mater Foundation of the University of Bologna a prototype Plasma Focus, named PFMA-3 (for: ***Plasma Focus for Medical Applications number 3***), has been designed, built and put into operation.

PFMA-3 is being optimized as an X-ray generator, in which X-rays are obtained from the conversion of the electron beam mentioned above, driven through an extraction guide to impinge on a 50 μm brass foil. Throughout the experimental campaign presented here, the main working parameters were invariably kept as follows: charging voltage: 18 kV; working gas: nitrogen; gas pressure: 45 Pa; bank capacitance: 22 μF; overall inductance: 150 nH; maximum current: 220 kA.

In view of applying the device to the irradiation of cells, which is the object of an impending radiobiological investigation, a method for fast and reliable determination of the dose delivered was sought. GAFCHROMIC® film (GF) was identified as a convenient detector, and aim of the present work was to design and test a procedure, based on GF, that would meet the expected requirements of speed and dependability.

**2. Gafcromic® Film**
As sensitive material in this campaign GF was utilized, of two kinds: EBT-2 (International Specialty Products - Advanced Materials Group a) and HD-810 (International Specialty Products - Advanced Materials Group b). This product has the two properties of increasing its optical thickness proportionally to the adsorbed dose and of being tissue equivalent. The product comes in sheets, from which pieces of the appropriate dimensions (henceforth referred to as a "samples") were cut. Care was taken to maintain the same orientation (with respect to the longitudinal direction of the original sheet) in reading all the samples, as recommended by the GF manufacturer. The procedure used throughout this campaign to read the films was as follows: samples were scanned on an HP LaserJet M1522 nf, at 300 dpi resolution, and the image converted to an 8-bit grey scale (256 shades of grey). Devic S *et al* (2004) have shown that GF can be scanned in reflection mode on a flatbed scanner. Every sample was first scanned before exposure to determine background optical density: the grey reading was averaged throughout the sample surface (i.e., over all the pixels) and a background value $Z_b$ obtained. The sample was measured after exposure, the reading averaged over the exposed surface and then the average background subtracted, thus generating the net reading $Z_n$.
To correlate the net reading to the absorbed dose calibration was necessary. As the two types of GF used have significantly different behaviours and were utilized in very different fashions, they will be discussed separately.

*2.1. EBT2 film*
For this type of GF the measurable dose range is 0.01÷8 Gy. As can be seen in figure 1, the composition is asymmetric on the two sides of the active layer.

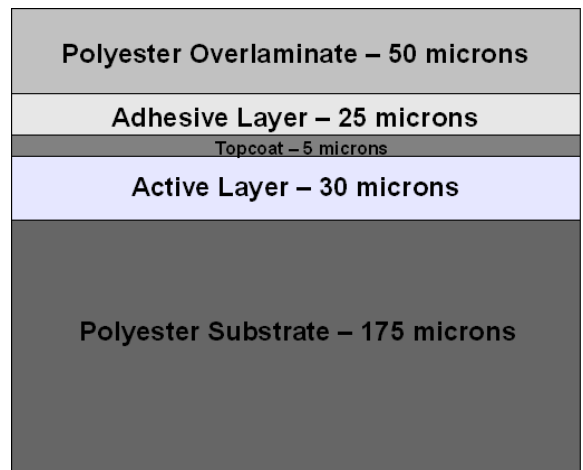

Figure 1: Gafchromic® film:EBT2

Details on the composing material for GF are presented in table 1.

**Table 1.** Mass fraction composition of GAFCHROMIC® film

|   | H | C | O | Li | Cl |
|---|---|---|---|---|---|
| Polyester | 0.042 | 0.628 | 0.335 | - | - |
| Active layer | 0.097 | 0.591 | 0.285 | 0.009 | 0.018 |
| Adhesive layer | 0.094 | 0.651 | 0.255 | - | - |
| Surface layer | 0.058 | 0.310 | 0.250 | 0.063 | 0.320 |

*2.1.1 Energy response*
The producer rates the response to photons of this type of GF as energy independent throughout the range 50÷1000 keV, however its useful range can be extended to lower energies, as reported in Arjomandy B *et al* (2010). For a further confirmation of this fact, two samples were irradiated with the same 1 Gy dose at two different photon energies. The two irradiations were conducted at the Comecer calibration center (Centro di

EBT2 Dosimetry of X-rays produced by the electron beam from PFMA-3

Taratura - COMECER SIT n.065/r Castel Bolognese, Italy), with a BALTEAU CSC320/70 X-ray tube. The ISO-4037 specified H-30 and H-60 beams (ISO-4037-1 1998) were selected. The working parameters of the H-30 beam are as follows: voltage: 30 kV; current: 14 mA; additional aluminum filter: 0.52 mm. These conditions produce a continuous spectrum with a mean energy of 19.7 keV. The working parameters for the H-60 beam are: voltage: 60kV; current: 11mA; additional aluminum filter: 3.2 mm. The continuous spectrum produced has a mean energy of 37.3 keV.
The results, in terms of levels of grey, are presented in table 2.

**Table 2.** Sensitivity at low energy for EBT2 film

| Beam | Energy [keV] | $Z_b$ | $Z_n$ | Dose [Gy] |
|---|---|---|---|---|
| H-30 | 19.7 | 60 | 64.6 | 1.01 |
| H-60 | 37.3 | 60.8 | 67.5 | 1.00 |

As before, $Z_b$ is the background grey level and $Z_n$ the net grey level after subtraction of the background value. The resulting difference is of the order of 4%, confirming further that GF can be utilized at the energies, lower than 50 keV, of interest to the present paper.

*2.1.2 Dose calibration*
The response of EBT2 GF is not linear with dose, therefore a number of experimental points is needed to determine a calibration curve. To this end, measurements were performed with the Philips MG323 X-ray tube operated by the Radiation protection Institute of ENEA in Bologna (Centro di taratura - Istituto di Radioprotezione, ENEA, Bologna, Italy).
The beam utilized is the above mentioned H-30, as the spectrum emitted is quite comparable to the one produced by PFMA-3. Twenty samples of size 2x2 cm, were positioned on a Plexiglas holder (figure 2) at 60 cm from the focus of the tube, and were given 20 different dose values.

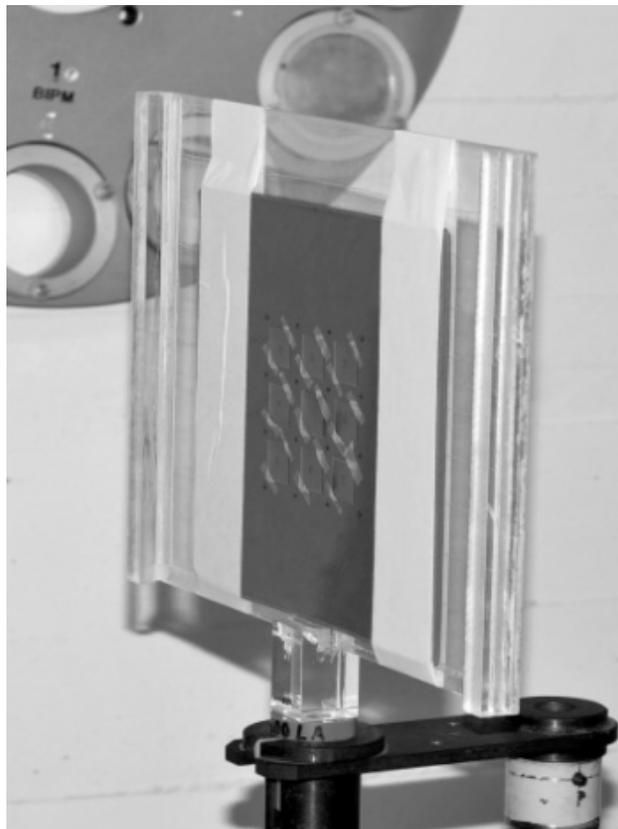

Figure 2: Plexiglas support for the irradiation of the gafchromic film

The results are presented in table 3, where $Z_b$ and $Z_n$ have the meaning already discussed, whereas Std_$Z_b$ is the standard deviation calculated from the readings of all the pixels in the exposed surface.

**Table 3.** Grey values in the irradiate samples

| Dose [Gy] | $Z_b$ | $Z_n$ | $Std\_Z_b$ |
|---|---|---|---|
| 0.05 | 55.9 | 6.9 | 0.19 |
| 0.1 | 56.1 | 15.6 | 0.26 |
| 0.3 | 58.7 | 29.7 | 0.21 |
| 0.5 | 56.1 | 44.7 | 0.31 |
| 0.6 | 54.2 | 46.6 | 0.35 |
| 0.7 | 55.8 | 53.2 | 0.09 |
| 1 | 56.9 | 64.7 | 0.23 |
| 1.3 | 56.2 | 76 | 0.36 |
| 1.5 | 55.3 | 84.4 | 0.37 |
| 1.7 | 56.7 | 88.6 | 0.21 |
| 2 | 55.1 | 98.1 | 0.18 |
| 2.5 | 56.3 | 105.5 | 0.32 |
| 3 | 56.1 | 115 | 0.38 |
| 3.5 | 55.3 | 122.3 | 0.34 |
| 4 | 55.7 | 127.6 | 0.27 |
| 4.5 | 55.1 | 132.4 | 0.29 |
| 5 | 54.6 | 136.9 | 0.39 |
| 6 | 55.9 | 144.1 | 0.25 |
| 7 | 55 | 149 | 0.33 |
| 8 | 54.5 | 150 | 0.19 |

It was observed that doses of 7 and 8 Gy produce grey levels that are too close. Therefore, EBT2 film will not be used in the present investigation for doses above 7 Gy.

The calibration curve, shown in figure 3, is obtained from the 20 experimental values fitted with a 4$^{th}$ order polynomial. At each value is associated an error bar of ± 2.5% on the Y-axis, which is the reported uncertainty in the dose delivered by the X-ray tube.

The quantity on the horizontal axis is the net grey level value after subtraction of the background. The correlation factor ($R^2$=0.9995) and the scatter plot, figure 4, indicate a very good fitting.

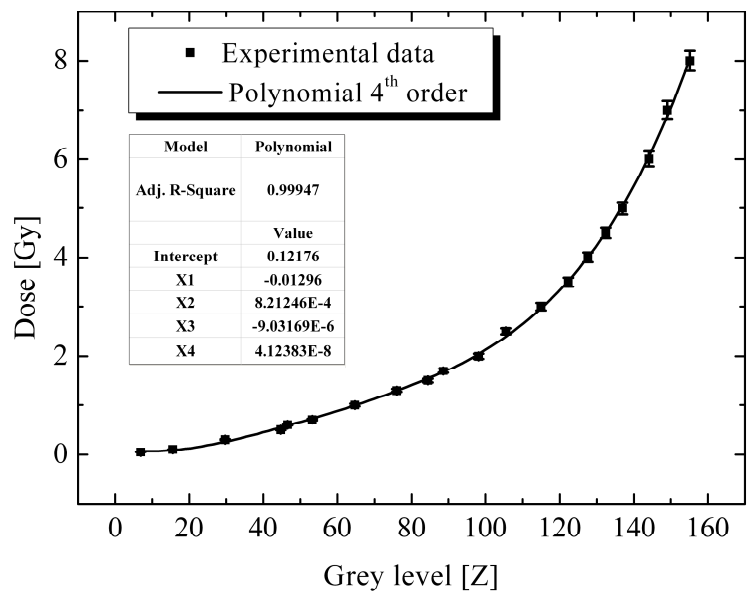

Figure 3: Calibration curve

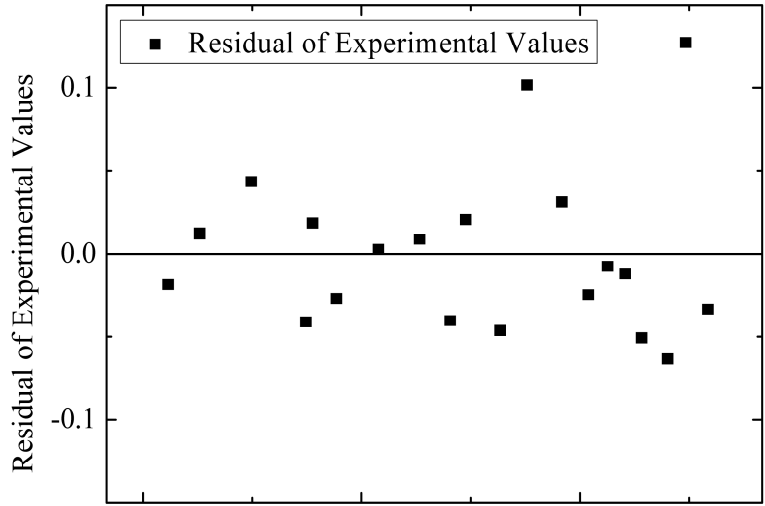

Figure 4: Residuals of the experimental values

*2.2. HD-810 film*

HD-810 type film is designed to measure absorbed dose from high-energy photons or electrons. The response of this film is energy-independent for photons above 0.2 MeV and is linear with dose. The rated dose range is 10÷250 Gy. Figure 5 shows the structure of this film.

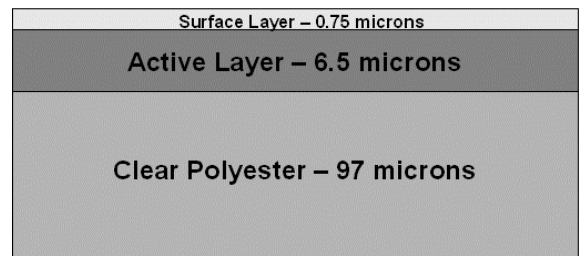

Figure 5: Gafchromic film: HD-810

In the present work, however, this GF is used with a different purpose: since some of the most energetic electrons cross the brass foil and impinge on the dosimetric material, it was found appropriate to interpose a filter, ideally one that would stop all the electrons without affecting the X-rays. On the other hand, the discharges cannot be all consistently identical, and a fast, even though approximate means to judge the quality of the discharges as far as radiation production would be useful.

HD-810 GF meets both these needs: it stops all stray electrons attenuating only slightly the X-ray beam; and it affords a qualitative indication of the intensity of the mixed radiation field (electrons and X-rays) impinging on it. Since the film never saturates with the doses delivered in the present experiments, the intensity can be appraised visually by the operator from the degree of darkening of the film. Again, this appraisal cannot be but qualitative, and in relative terms: stronger or weaker discharges; yet it proves useful. No quantitative information can be drawn, given that the X-rays present have energies far below the recommended range, and anyhow the electron component is also present. Under these circumstances there is no question of calibrating this type of GF.

**3. Cell culture irradiation**

For the radiobiological investigation to come, electrons produced in the pinch phase will be directed onto the brass target to produce X-rays, which will in turn irradiate a cell culture contained in a holder specifically designed, see figure 6, that attaches at the end of the electron extraction tube, after the brass foil.

EBT2 Dosimetry of X-rays produced by the electron beam from PFMA-3

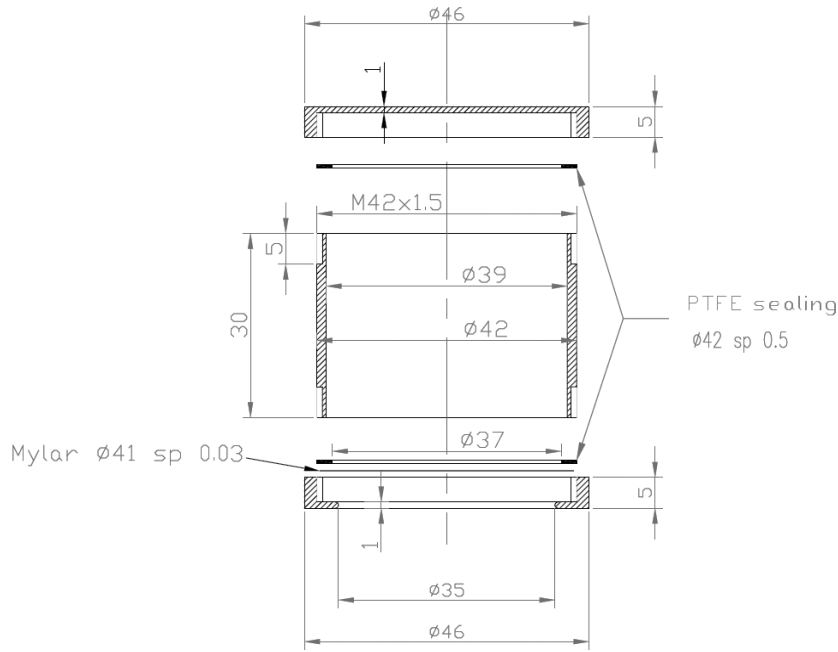

Figure 6: cell holder

The holder is comprised of a hollow, cylindrical central piece 3 cm long and with a 3.9 cm inner diameter, threaded at both ends to accommodate sealing end pieces. The end piece toward the brass converter has a 3.5 cm diameter aperture, and is sealed with a 30 μm Mylar® foil seated between the central body and the end piece. The cell culture is contained inside the holder, resting on the Mylar® foil.

To assess the dose delivered to the cells a stack of GF was inserted between the brass foil and the cell holder. The idea was to measure the dose to the last GF layer (the one immediately adjacent to the Mylar® foil of the holder) and from that infer the dose to the cell layer. To this end the cell layer was simulated with a further GF layer (the absorption in the Mylar® foil was neglected, this point will be addressed in detail shortly): so the layer could be added to the stack, this latter irradiated, and the doses to the GF layers measured. Repeating the experiment a sufficient number of times, the relationship between the dose to the extra layer and the dose to the layer immediately preceding could be determined. Hence, determination of this latter dose will serve to appraise the dose to the cells.

*3.1. Dosimetric Stack*

Initially, a stack of 5 GF EBT2 layers was used, as presented in figure 7, and series of 5 discharges were loaded on the stack.

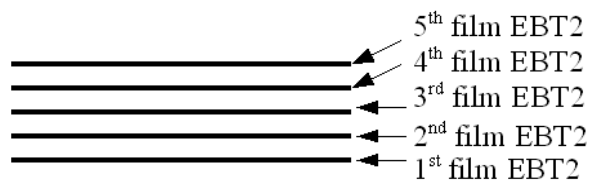

Figure 7: 5-layer, EBT2 stack

However, for the reasons discussed in 2.2 and because of the high doses present, the first two films were completely saturated, and occasionally the third as well. It was chosen to limit the number of discharges per run to only 4 and to use as first layer the HD-810 film. This configuration is shown in figure 8.

EBT2 Dosimetry of X-rays produced by the electron beam from PFMA-3

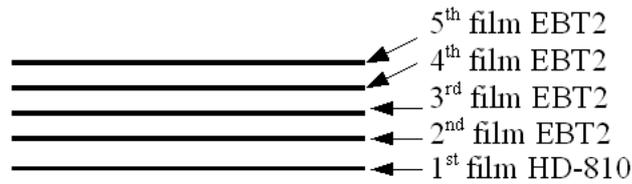

Figure 8: mixed-layer, HD-810 and EBT2 stack

Both HD-810 and EBT2 films are all consistently positioned with the thinner coating (0.75/0.80 μm) facing toward the incoming beam. The sample area is 4.5×4.5 cm. Figure 9 presents an example of GF after irradiation.

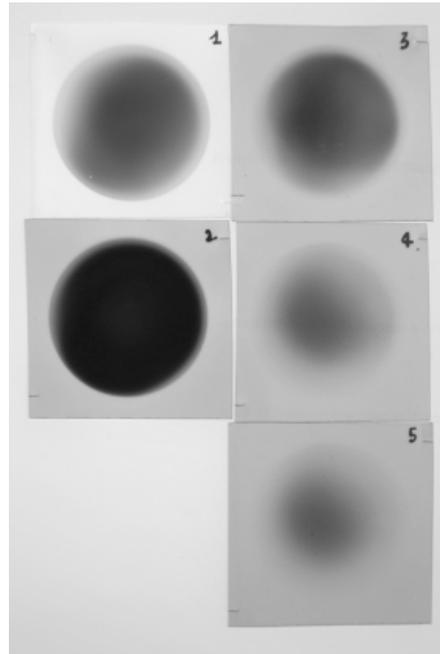

Figure 9: Irradiation of 5 gafchromic films

After the average level of grey $Z_n$ in the most exposed area was determined, with the calibration curve the $Z_n$ values were converted to doses. Table 4 is an example of the results.

Table 4. Results of one series of 4 discharges

|  | 1st film (HD-810) | 2nd film (EBT2) | 3rd film (EBT2) | 4th film (EBT2) | 5th film (EBT2) |
|---|---|---|---|---|---|
| $Z_b$ | 1 | 88 | 90 | 89 | 90 |
| $Z_n$ | 130 | 150 | 116 | 103 | 90 |
| Dose [Gy] |  | 7.1 | 3.0 | 2.2 | 1.8 |

Dose values for the first layer could not be meaningfully reported as discussed in 2.2.
The experimental campaign was conducted running 5 series of 9 repeated measurements (each of them counting 4 discharges), and a correlation between the doses to the 4th and 3rd film, as well as that between the doses to the 5th and 4th film, was determined. Figure 10 presents the results of one such series (9 measurements), reporting the ratio between doses to the the 4th and 3rd film and that between the doses to the 5th and 4th film.

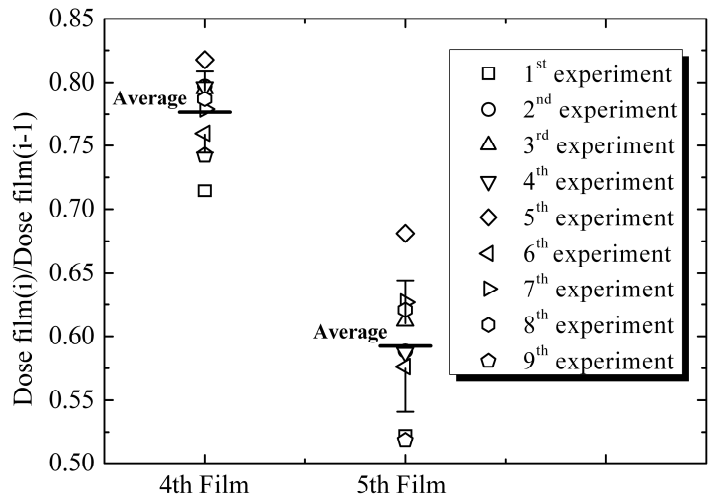

Figure 10: First measurement campaign

It can be seen that the ratio of doses of films 4th to 3rd presents much less fluctuation, and therefore it was chosen as the best configuration: one HD-810 film followed by two EBT2 films, then the cell holder. The same pattern was observed in all the measurements of the 5 series, and the overall average ratio between the doses to the $3^{rd}$ and $4^{th}$ film was calculated at 0.78±0.03. Therefore the dose is read from the $3^{rd}$ film, and multiplied by the average ratio to estimate the dose to the cell culture.

*3.2. Attenuation in the Mylar® foil*
It remains to be ascertained that the attenuation in the Mylar® foil is indeed negligible. Figure 11 (Center for X-Ray Optics 2010) shows the transmission coefficient for varying thicknesses of Mylar®, as a function of photon energy. Of interest here the curve for 30 μm.

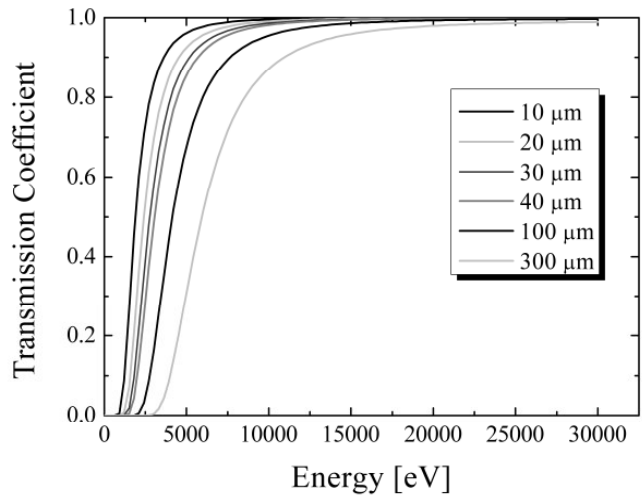

Figure 11: Transmission coefficient for Mylar® foils of varying thicknesses

To investigate the attenuation in the Mylar® foil, two sets of experiments were conducted, with two different film stack schemes.
In the first set the initial stack was used, 5 EBT2-type GF, to which a Mylar® foil disc was added between the $1^{st}$ and the $2^{nd}$ film, see figure 12.

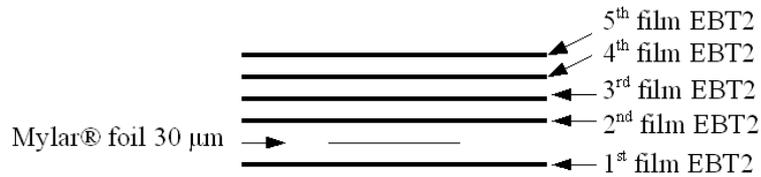

Figure 12: First configuration: foil between the first and second film

Figure 13 shows quite clearly the difference in the grey level of the second film between the area covered by the foil and the rest of the exposed area.

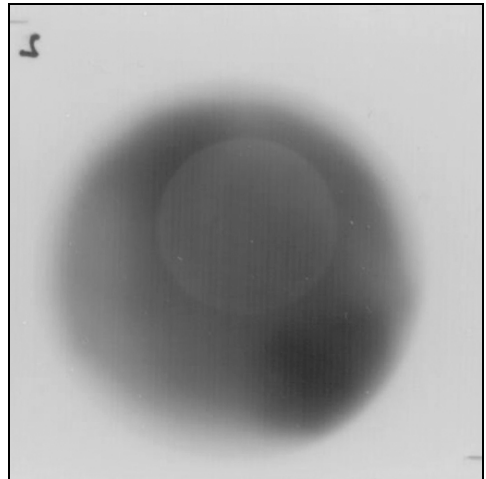

Figure 13: Second gafchromic film

The grey level values were averaged in the area covered by the disc and in the rest of the exposed area and the background subtracted from both values: the resulting $Z_n$ is shown in table 5, confirming the difference detected visually.

**Table 5.** Effect of attenuation in Mylar®

| $Z_n$ within covered area | $Z_n$ outside covered area |
|---|---|
| 148.1 | 153.4 |

It may be noted incidentally that since the above values show attenuation in the 30 μm Mylar® foil, they confirm the existence of a component below 30 keV in the photon spectrum.
The second set of measurements was conducted using the final stack, as described in the previous paragraph, in which the same Mylar® foil disc was inserted between the 3rd and the 4th film, see figure 14.

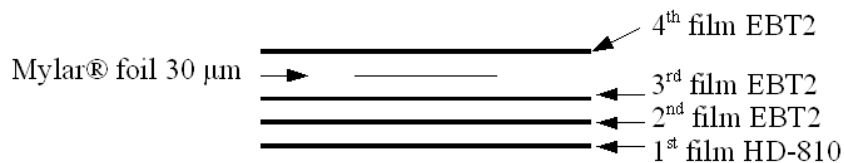

Figure 14: Second configuration: foil between third and fourth film

No variations in grey scale due to the presence of the foil was detected, or in other words, the transmission in Mylar® was 100%. By the way, from Figure 11 it is inferred that photons capable of reaching the Mylar® foil have energies above 30 keV.

## 5. Conclusions
As mentioned already, it is intended to conduct a radiobiological experimental campaign to assess the potential of the Plasma Focus as a source of X-ray for IORT. In this campaign, X-rays from PFMA-3 will be directed to cell cultures, prepared in a cell holder designed to this effect. The relationship between doses delivered by PFMA-3 and radiobiological effect on cells will be investigated. To this end, a fast and reliable method for assessing the doses imparted to the cell cultures has been devised and characterised. The method relies on the use of Gafchromic® film, a well consolidated sensitive medium, and on a procedure specifically designed. It is expected that it will yield fast yet accurate results.

**Acknowledgements**
Work performed in the framework of a Research Contract of Alma Mater s.r.l. of the University of Bologna and with the financial support of ABO Foundation.